\pdfoutput=1
\documentclass[12pt,a4paper]{article}
\usepackage{lineno}  % for line numbering during review

\usepackage{graphicx}  % to include figures (can also use other packages)

\usepackage{xspace}
\usepackage{color}
\usepackage{colortbl}
\usepackage{subfigure}
\usepackage{amsmath}

\usepackage{ifthen} % for conditional statements
\newboolean{pdflatex}
\setboolean{pdflatex}{true} % use this if using non-eps figures
\usepackage{rotating} 
\newboolean{articletitles}
\setboolean{articletitles}{true} 

\newboolean{uprightparticles}
\setboolean{uprightparticles}{false} %Set to true to get roman particle symbols
\usepackage{amssymb}
\usepackage{amsfonts}
\usepackage{upgreek}
\usepackage{lhcb-symbols-def}
\usepackage{hyperref}
\usepackage[all]{hypcap} 

\usepackage{bm}
\usepackage{afterpage}

\newcommand{\bea}{\begin{eqnarray}}
\newcommand{\eea}{\end{eqnarray}}
\newcommand{\beq}{\begin{equation}}
\newcommand{\eeq}{\end{equation}}

\def\BsbbarParen {\ensuremath{\stackrel{(\rule[0.2em]{0.5em}{0.01em})}{\Bs}}}

\usepackage{mciteplus}

\begin{document}

%%%%%%%%%%%%%%%%%%%%%%%%%
%%%%% Title     %%%%%%%%%
%%%%%%%%%%%%%%%%%%%%%%%%%

% %%%%%%% CHOOSE --------
%  Choose the right title template or customize existing one if your type is still missing

% $Id: titlepage.tex 2394 2011-02-17 10:11:10Z gcowan $
% ===============================================================================
% Purpose: LHCb-ANA Note title page template
% Author: 
% Created on: 2010-10-05
% ===============================================================================

%%%%%%%%%%%%%%%%%%%%%%%%%
%%%%  TITLE PAGE  %%%%%%
%%%%%%%%%%%%%%%%%%%%%%%%%
\begin{titlepage}
%Primary authors Bilas Pal, Sheldon Stone and Liming Zhang
% Header ---------------------------------------------------
\belowpdfbookmark{Title page}{title}

\pagenumbering{roman}
\vspace*{-1.5cm}
\centerline{\large EUROPEAN ORGANIZATION FOR NUCLEAR RESEARCH (CERN)}
\vspace*{1.5cm}
\hspace*{-5mm}\begin{tabular*}{16cm}{lc@{\extracolsep{\fill}}r}
\vspace*{-12mm}\mbox{\!\!\!\includegraphics[width=.12\textwidth]{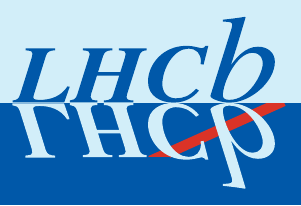}}& & \\
&& CERN-PH-EP-2011-205\\
&& LHCb-PAPER-2011-031\\
&&December 13, 2011 \\
\end{tabular*}
\vspace*{4cm}
\begin{center}

{\bf\huge\boldmath 
Measurement of the \CP violating phase $\phi_s$ in $\Bsb\to J/\psi f_0(980)$
}\\
\vspace*{2cm}
\normalsize {
The LHCb Collaboration\footnote{Authors are listed on the following pages.}
%========================================================================%
}
\end{center}

% Abstract -----------------------------------------------
\begin{abstract}
  \noindent
Measurement of mixing-induced \CP violation in $\Bsb$ decays is of prime importance in probing new physics. So far only the channel $\Bsb\to J/\psi\phi$ has been used. Here we report on a measurement using an LHCb data sample of 0.41\,fb$^{-1}$, in the \CP odd eigenstate $J/\psi f_0(980)$, where $f_0(980)\to\pi^+\pi^-$. A time dependent fit of the data with the $\Bsb$ lifetime and the difference in widths of the heavy and light eigenstates constrained to the values obtained from  $\Bsb \to J/\psi\phi$ yields a value of the \CP violating phase of $-0.44\pm 0.44\pm0.02~{\rm\,rad}$, consistent with the Standard Model expectation.

\end{abstract}

\vspace*{2.0cm}
\hspace*{6mm}{\it PACS:} 13.25.Hw, 14.40.Nd, 11.30.Er\\
\hspace*{6mm}Submitted to Physics Letters B\\
\newpage
% Authors -------------------------------------------------
\begin{center}
The LHCb Collaboration\\
\begin{flushleft}
%{\Large LHCb Collaboration ----- official authorship list}\\[4ex]
%valid for date: 3. Nov. 2011\\
%used for paper: Measurement of $\phi_s$ in $B_s \to J/\psi f_0(980)$ (LHCb-PAPER-2011-031)\\[4ex]
%collaborators included, who did not leave before 3. Nov. 2010\\
%                           and who joined before 3. May. 2011\\[2ex]
%{\small today is 4. Dec. 2011}\\[4ex]
%-- 
%-- LHCb Authorlist, Status of 3. Nov. 2011
%-- 
R.~Aaij$^{23}$, 
C.~Abellan~Beteta$^{35,n}$, 
B.~Adeva$^{36}$, 
M.~Adinolfi$^{42}$, 
C.~Adrover$^{6}$, 
A.~Affolder$^{48}$, 
Z.~Ajaltouni$^{5}$, 
J.~Albrecht$^{37}$, 
F.~Alessio$^{37}$, 
M.~Alexander$^{47}$
% ...

{\footnotesize \it
$ ^{1}$Centro Brasileiro de Pesquisas F\'{i}sicas (CBPF), Rio de Janeiro, Brazil\\
% ...
$ ^{5}$Clermont Universit\'{e}, Universit\'{e} Blaise Pascal, CNRS/IN2P3, LPC, Clermont-Ferrand, France\\
$ ^{6}$CPPM, Aix-Marseille Universit\'{e}, CNRS/IN2P3, Marseille, France\\
% ...
$ ^{23}$Nikhef National Institute for Subatomic Physics, Amsterdam, The Netherlands\\
% ...
$ ^{35}$Universitat de Barcelona, Barcelona, Spain\\
$ ^{36}$Universidad de Santiago de Compostela, Santiago de Compostela, Spain\\
$ ^{37}$European Organization for Nuclear Research (CERN), Geneva, Switzerland\\
% ...
$ ^{42}$H.H. Wills Physics Laboratory, University of Bristol, Bristol, United Kingdom\\
% ...
$ ^{47}$School of Physics and Astronomy, University of Glasgow, Glasgow, United Kingdom\\
$ ^{48}$Oliver Lodge Laboratory, University of Liverpool, Liverpool, United Kingdom\\
% ...
$ ^{n}$LIFAELS, La Salle, Universitat Ramon Llull, Barcelona, Spain\\
% ...
}
\bigskip
%---- LHCb Authorlist, Status 3. Nov. 2011
%---- Number of Authors = 587

\end{flushleft}
\end{center}
%%\vspace*{2.0cm}
%%\vspace{\fill}

\end{titlepage}
%\tableofcontents
\newpage

\pagestyle{empty}  % no page number for the title 

%%%%%%%%%%%%%%%%%%%%%%%%%%%%%%%%
%%%%%  EOD OF TITLE PAGE  %%%%%%
%%%%%%%%%%%%%%%%%%%%%%%%%%%%%%%%

%  empty page follows the title page ----

%\newpage
%\setcounter{page}{2}
%\mbox{~}
%

%\input{title-LHCb-CONF}
%\input{title-LHCb-PAPER}
% %%%%%%%%%%%%% ---------

%%%%%%%%%%%%%%%%%%%%%%%%%%%%%%%%
%%%%%  Table of Content   %%%%%%
%%%%%%%%%%%%%%%%%%%%%%%%%%%%%%%%
%%%% Uncomment next 2 lines if desired

%\tableofcontents

%%%%%%%%%%%%%%%%%%%%%%%%%
%%%%% Main text %%%%%%%%%
%%%%%%%%%%%%%%%%%%%%%%%%%

\pagestyle{plain} % restore page numbers for the main text
\setcounter{page}{1}
\pagenumbering{arabic}

% %%%%%%% CHOOSE --------
%% ----------------------------------
%% Line numbering on the left margin 
%% ----------------------------------
%% Uncomment during review phase. 
%% Comment it out before a final submission.
%\linenumbers
%% --------------------------------
% %%%%%%%%%%%%% ---------

%  You can include short sections directly in the main tex file. 
%  However, for larger papers it is desirable to split 
%  the text into several semiautonomous files, which can be revised independently. 
%  This is especially useful when developing a document in collaboration with several people, 
%  since then different parts can be edited independently. 
%  This type of file organization is shown here. 
% 

%\clearpage
%{\noindent\bf\Large Appendix}
%\appendix
%\input{appendixA}

%\addcontentsline{toc}{section}{References}

\section{Introduction}

\begin{figure}[hbt]
\centering
\includegraphics[width=3.5in]{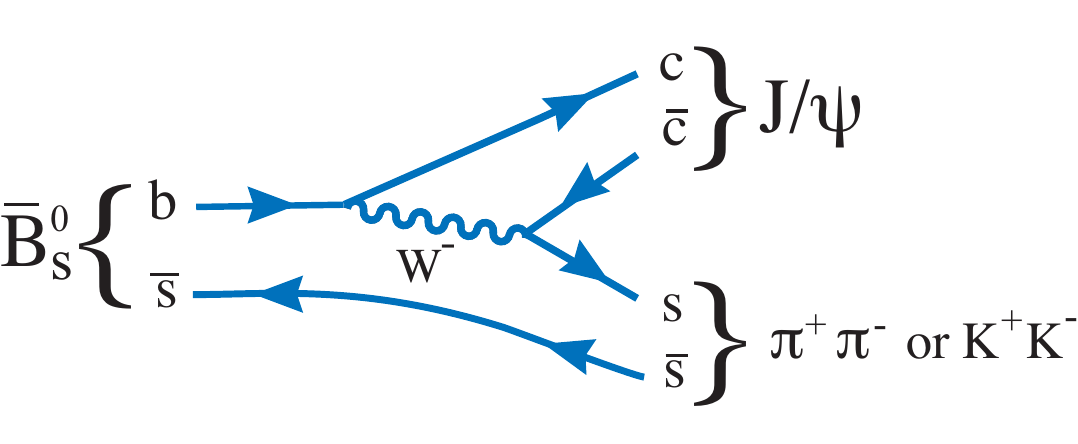}
\caption{Dominant decay diagrams for $\Bsb\to J/\psi f_0(980)$ or $J/\psi\phi$ decays.
 } \label{psi_f0-phi}
\end{figure}

An important goal of heavy flavour experiments is to measure the mixing-induced
\CP violation phase in $\Bsb$ decays, $\phi_s$. As this phase is predicted to be small in the Standard Model (SM) \cite{Charles:2011va}, new physics can induce large changes \cite{Dunietz:2000cr}. Here we use the decay mode $\Bsb\to J/\psi f_0(980)$. If only the dominant decay diagrams shown in  Fig.~\ref{psi_f0-phi} contribute, then the value of $\phi_s$ using $\Bsb\to J/\psi f_0(980)$ is the same 
as that measured using $\Bsb\to J/\psi \phi$ decay. 

Motivated by a prediction in Ref. \cite{Stone:2008ak}, LHCb searched for and made the first observation of $\Bsb\to J/\psi f_0(980)$ decays \cite{Aaij:2011fx}  that was subsequently confirmed by other experiments \cite{Li:2011pg,*Abazov:2011hv,Aaltonen:2011nk}. 
Time dependent \CP violation can be measured without an angular analysis, as the final state is a \CP eigenstate. From now on $f_0$ will stand only for $f_0(980)$. 

 In the Standard Model, in terms of CKM matrix elements, $\phi_s= -2\arg\left[ \frac{V_{ts}V_{tb}^*}{V_{cs}V_{cb}^*}\right].$  The equations below are written assuming that there is only one decay amplitude, ignoring  possible small contributions from other diagrams \cite{Faller:2008gt,*Fleischer:2011au}.
The decay time evolutions for initial $B_s^0$ and $\overline{B}_s^0$ are \cite{Nierste:2009wg,*Bigi:2000yz}
\begin{eqnarray}
   \label{eq:CPrate}
\Gamma\left(
%(\stackrel{_{({\rm\bold\Huge -})}}{B^0_s}
\BsbbarParen\to J/\psi f_0\right) &=& {\cal N}e^{-\Gamma_s t}\, \Bigg\{e^{\Delta\Gamma_s t/2}(1+\cos\phi_s)+ e^{-\Delta\Gamma_s t/2}(1-\cos\phi_s) \nonumber\\
&&~~~~~~~~~~~~~~ \pm\sin\phi_s\sin \left( \dm_s \, t \right) \Bigg\}, 
\end{eqnarray}
where $\Delta\Gamma_s$ is the decay width difference between light and heavy mass eigenstates, $\Delta\Gamma_s=\Gamma_{\rm L}-\Gamma_{\rm H}$. The decay width $\Gamma_s$ is the average of the widths $\Gamma_{\rm L}$ and $\Gamma_{\rm H}$, and $ {\cal N}$ is a time-independent normalization factor. The plus sign in front of the $\sin\phi_s$ term applies to an initial $\overline{B}^0_s$ and the minus sign for an initial $B^0_s$ meson.
The time evolution of the untagged rate is then
\begin{equation}
\Gamma\left( B_s^0\to J/\psi f_0\right) +  \Gamma\left( \Bsb\to J/\psi f_0\right)=
{ \cal N}e^{-\Gamma_s t}\, \Bigg\{e^{\Delta\Gamma_s t/2}(1+\cos\phi_s)+ e^{-\Delta\Gamma_s t/2}(1-\cos\phi_s)\Bigg\}.
\label{eq:untagged}
\end{equation}
Note that there is information in the shape of the lifetime distribution that correlates $\Delta\Gamma_s$ and $\phi_s$.
In this analysis we will use both samples of flavour tagged and untagged decays. Both Eqs.~\ref{eq:CPrate} and \ref{eq:untagged} are insensitive to the change $\phi_s\to \pi-\phi_s$ when $\Delta\Gamma_s\to -\Delta\Gamma_s$.

\section{Selection requirements}
\label{sec:2}
% The data sample contains approximately 35\,pb$^{-1}$ of integrated luminosity collected in 2010, and an additional 345\,pb$^{-1}$ collected in the first half of 2011. 
%Event selection requirements are based on previous Monte Carlo simulation studies as reported in Ref.~\cite{f0-cuts}.
We use a data sample of 0.41\,fb$^{-1}$ collected in 2010 and the first half of 2011 at a centre-of-mass energy of 7 TeV. This analysis is restricted to events accepted by a $J/\psi\to\mu^+\mu^-$ trigger.
%Subsequent selection criteria are applied that serve to reject background, yet preserve high efficiencies, based on Monte Carlo simulations.
 The LHCb detector and the track reconstruction are  described in Ref.~\cite{LHCb-det}. The detector elements most important for this analysis are the VELO,  a silicon strip device that surrounds the $pp$ interaction region, and other tracking devices. Two Ring Imaging Cherenkov (RICH) detectors are used to identify charged hadrons, while muons are identified using their penetration through iron. 
%Chambers interspersed with iron are used to distinguish muons from hadrons.

To be considered
a $J/\psi\to\mu^+\mu^-$ candidate particles of opposite charge are required to have  transverse momentum, $p_{\rm T}$, greater than 500\,MeV, be identified as muons, and 
form a vertex with fit  $\chi^2$ per number of degrees of freedom (ndof) less than 11. We work in units where $c=\hbar=1$. Only candidates with dimuon invariant mass between $-$48~MeV to +43 MeV of the $J/\psi$ mass peak are selected. 
Pion candidates are selected if they are inconsistent with having been produced at the primary vertex. The impact parameter (IP) is the minimum distance of approach of the track with respect to the primary vertex. We require that the $\chi^2$ formed by using the hypothesis that the IP is zero be $>9$ for each track. For further consideration particles forming di-pion candidates must be positively identified in the RICH system, and must have their scalar sum $p_{\rm T}> 900$\,MeV.

To select $\Bsb$ candidates we further require that the two pions form a vertex with a $\chi^2< 10$, that they form a candidate $\Bsb$ vertex with the $J/\psi$ where the vertex fit $\chi^2$/ndof $<5$, that this vertex is $>1.5$\,mm from the primary, and  points to the primary vertex at an angle not different from its momentum direction by more than 11.8 mrad.

\begin{figure}[hbt]
\centering
\includegraphics[width=4.5in]{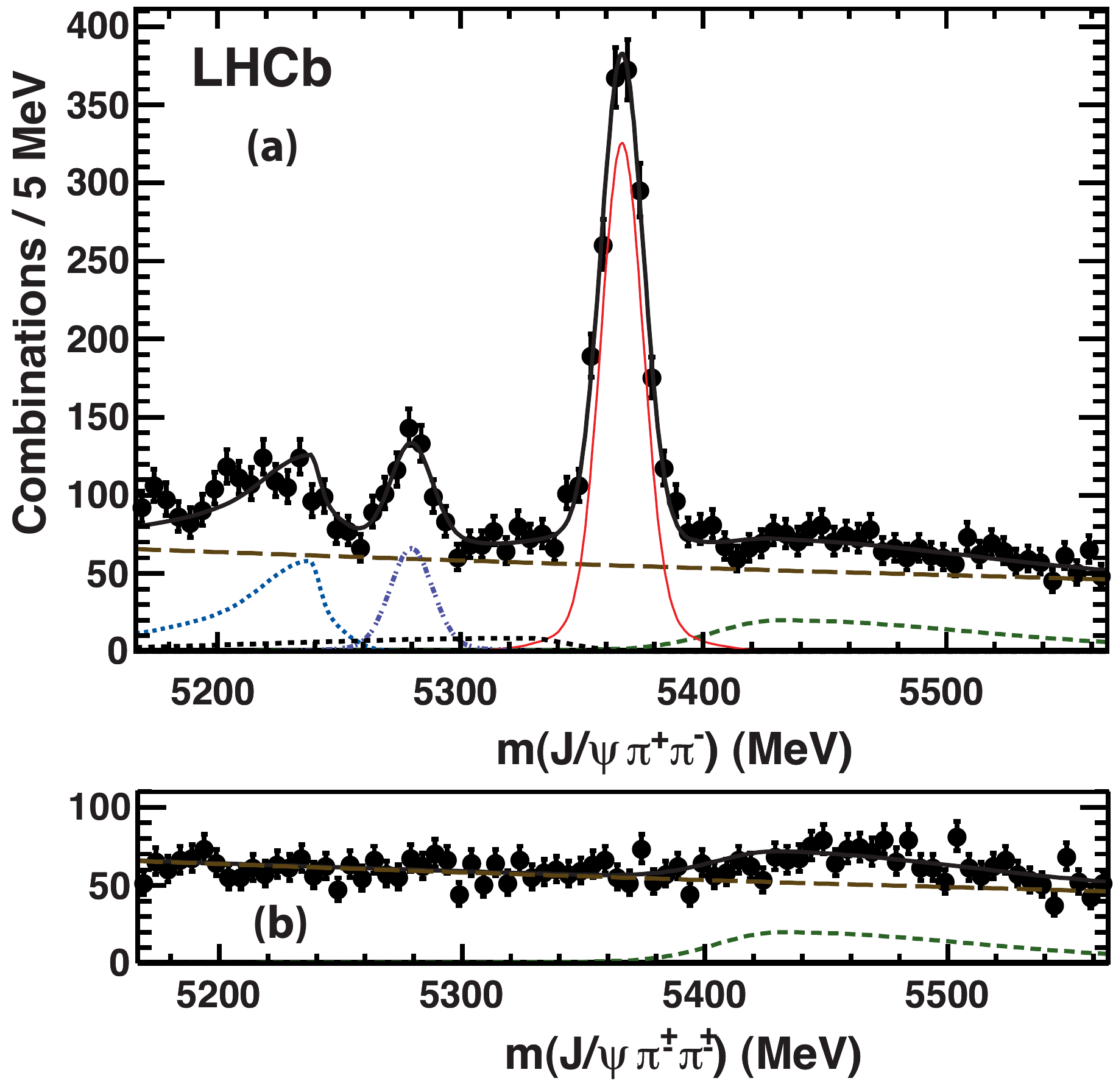}
\caption{(a) Invariant mass of $J/\psi\pi^+\pi^-$ combinations when the $\pi^+\pi^-$ pair is required to be within $\pm$90\,MeV of the nominal $f_0(980)$ mass. The data have been fitted with a double-Gaussian signal and several background functions. The thin (red) solid line shows the signal, the long-dashed (brown) line the combinatoric background, the dashed (green) line the $B^-$ background (mostly at masses above the signal peak), the dotted (blue) line the $\overline{B}^0\to J/\psi \overline{K}^{*0}$ background, the dash-dot line (purple) the $\overline{B}^0\to J/\psi \pi^+\pi^-$ background, the dotted line (black) the sum of $\Bsb\to J/\psi \eta'$ and $J/\psi\phi$ backgrounds (barely visible), and the thick-solid (black) line the total. (b) The mass distribution for like-sign candidates.} \label{fitmass_f0}
\end{figure}

The invariant mass of selected $\mu^+\mu^-\pi\pi$ combinations, where the di-muon pair is constrained to have the $J/\psi$ mass,
 is shown in Fig.~\ref{fitmass_f0} for both opposite-sign and like-sign di-pion combinations, requiring di-pion invariant masses within 90\,MeV of 980\,MeV. Here like-sign combinations are defined as the sum of $\pi^+\pi^+$ and $\pi^-\pi^-$ candidates. The signal shape, the same for both $\Bsb$ and $\overline{B}^0$,  is a double-Gaussian, where the core Gaussian's mean and width are allowed to vary, and the fraction and width ratio for the second Gaussian are fixed to the values obtained in a separate fit to $\Bsb\to J/\psi\phi$. The mean values of both Gaussians are required to be the same.
 The combinatoric background is described by an exponential function. Other background components are  $B^-\to J/\psi h^-$, where $h^-$ can be either a $K^-$ or a $\pi^-$ and an additional $\pi^+$ is found, $\Bsb\to J/\psi \eta'$, $\eta'\to \rho\gamma$, $\Bsb\to J/\psi \phi$, $\phi\to \pi^+\pi^-\pi^0$, and $\overline{B}^0\to J/\psi \overline{K}^{*0}$.  The shapes for these background sources are taken from Monte Carlo simulation based on PYTHIA \cite{Sjostrand:2006za} and GEANT-4 \cite{Agostinelli:2002hh} with their normalizations allowed to vary. We performed a simultaneous fit to the opposite-sign and like-sign di-pion
event distributions. There are 1428$\pm$47 signal events within $\pm$20\,MeV of the $\Bsb$ mass peak. The background under the peak in this interval is 467$\pm$11 events, giving a signal purity of 75\%.
Importantly, the like-sign di-pion yield at masses higher than the $\Bsb$ gives an excellent description of the shape and level of the background. Simulation studies have demonstrated that it also describes the background under the peak.

\begin{figure}[!hbt]
\centering
\includegraphics[width=4.5in]{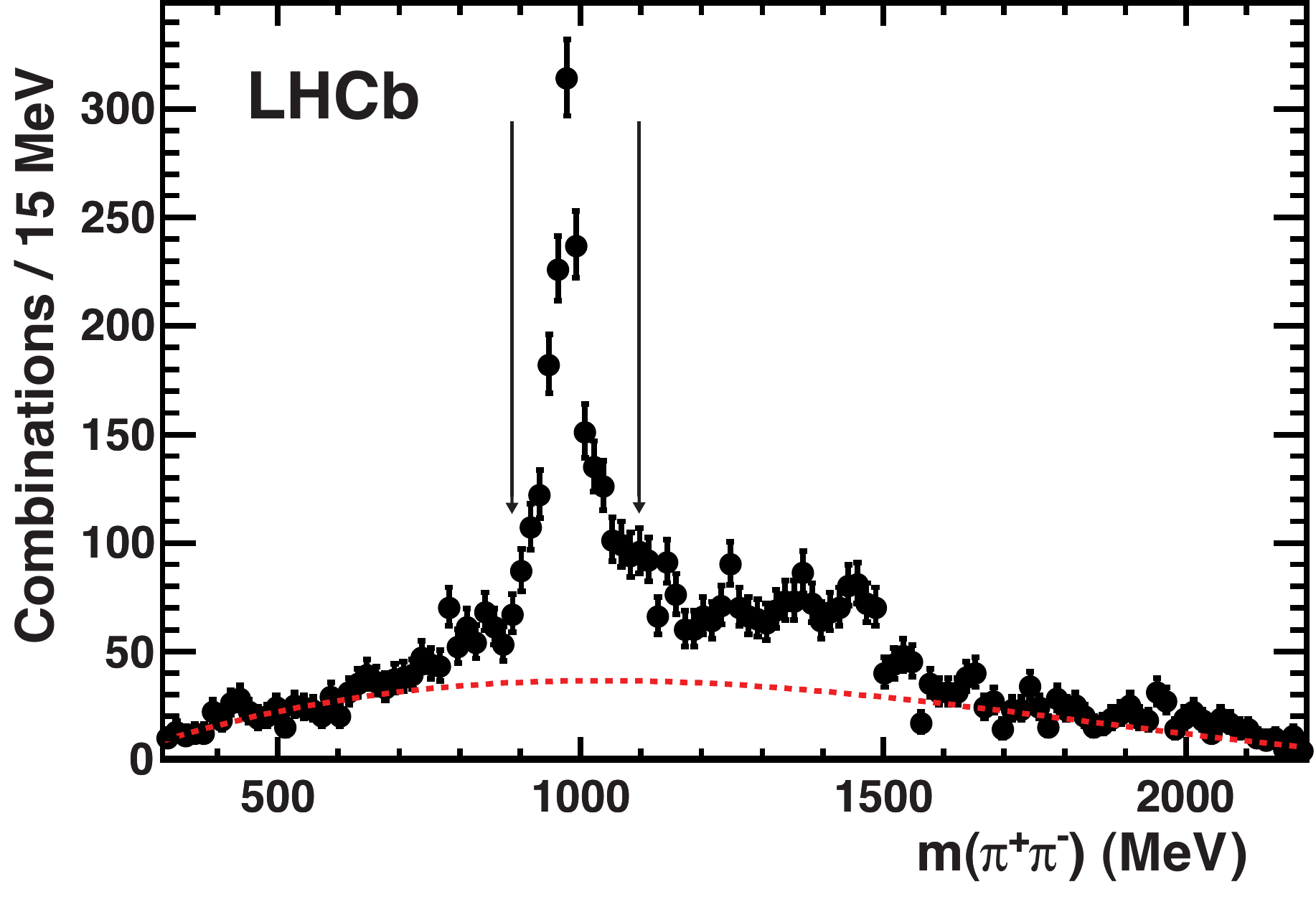}
\caption{Invariant mass of $\pi^+\pi^-$ combinations (points) and a fit to the  $\pi^{\pm}\pi^{\pm}$ data (dashed line) for events in the $\Bsb$ signal region. The region between the vertical arrows contains the events selected for further analysis. 
 } \label{mpipi}
\end{figure}

The invariant mass of di-pion combinations is shown in Fig.~\ref{mpipi} for both opposite-sign and like-sign di-pion combinations within $\pm$20\,MeV of the $\Bsb$ candidate mass peak. A large signal is present near the nominal $f_0(980)$ mass. Other $\Bsb\to J/\psi\pi^+\pi^-$ signal events are present at higher masses.  In what follows we only use events in the $f_0$ signal region from 890 to 1070\,MeV. 

\section{S-wave content}

Since the initial isospin of the $\s\overline{s}$ system that produces the two pions is zero, and since the $G$-parity of the two pions is even, only even spin is allowed for the $\pi^+\pi^-$ pair. Since no spin-4 resonances have been observed below 2 GeV,
the angular distributions are described by the coherent combination of spin-0 and spin-2 resonant decays. We use the helicity basis and define the decay angles as  $\theta_{J/\psi}$, the angle of the $\mu^+$ in the $J/\psi$ rest frame with respect to the $\Bsb$ direction, and $\theta_{f_0}$, the angle of the $\pi^+$ in the $\pi^+\pi^-$ rest frame with respect to the $\Bsb$ direction. 
The spin-0 amplitude is labeled as $A_{00}$, the three spin-2 amplitudes as $A_{2i}$, $i=-1,0,1$, and $\delta$ is the strong phase between the $A_{20}$ and $A_{00}$ amplitudes.

 After integrating over the angle between the two decay planes the joint angular distribution is given by \cite{Kutsckhe}
\begin{eqnarray}
\label{eq:amplitudes}
\frac{d\Gamma}{d\cos\theta_{f_0} d\cos\theta_{J/\psi}}&=&
\left| A_{00}+\frac{1}{2}A_{20}e^{i\delta}\sqrt{5}\left(3\cos^2\theta_{f_0}-1\right)\right|^2
\sin^2\theta_{J/\psi}\nonumber\\
&&+\frac{1}{4}\left(\left|A_{21}\right|^2+\left|A_{2-1}\right|^2\right)\left(15\sin^2\theta_{f_0}\cos^2\theta_{f_0}\right)
\left(1+\cos^2\theta_{J/\psi}\right).
\end{eqnarray}
Since the $\Bsb$ is spinless, when it decays into a spin-1 $J/\psi$ and a spin-0 $f_0$, $\theta_{J/\psi}$ should be distributed as $\sin^2\theta_{J/\psi}$ and  $\cos\theta_{f_0}$  should be uniformly distributed.  

\begin{figure}[hbt]
\centering
\includegraphics[width=6.in]{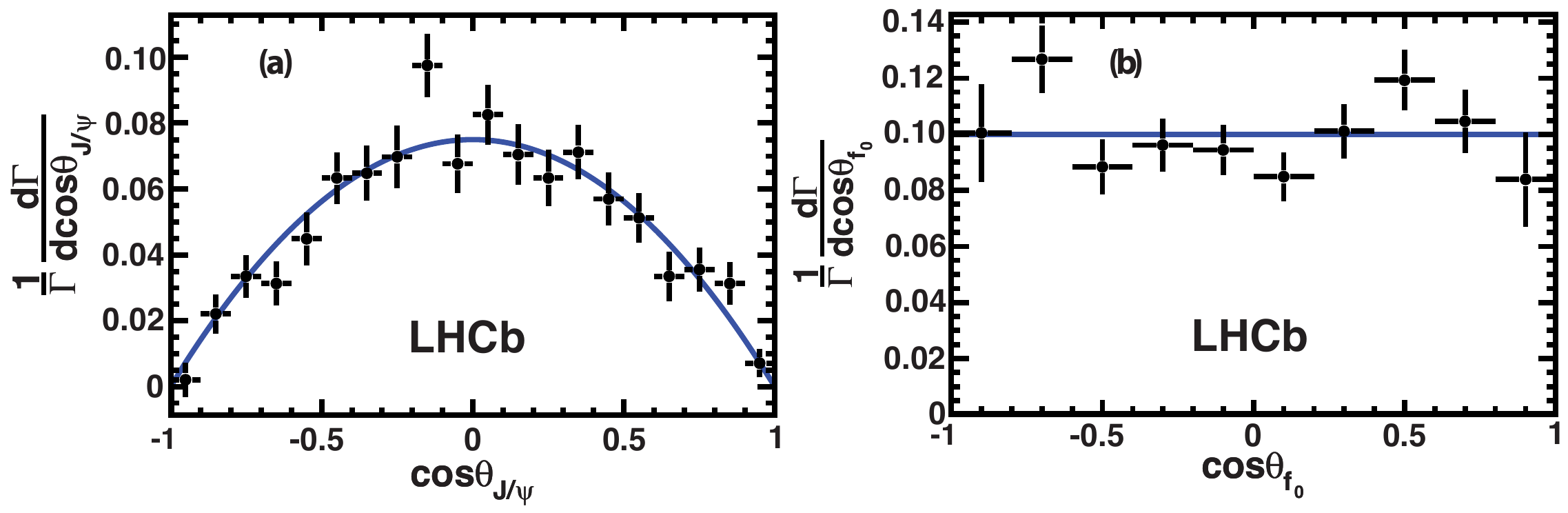}
\caption{Efficiency corrected, background subtracted angular distributions in the $\pi^+\pi^-$ mass region within $\pm$90\,MeV of 980\,MeV and within $\pm$20 MeV of the $\Bsb$ mass for (a) $\cos^2\theta_{J/\psi}$, and (b) $\cos\theta_{f_0}$.
The solid lines show the expectations for a spin-0 object.  
}
\label{hel-980}
\end{figure}

The helicity distributions of the opposite-sign data selected with reconstructed $J/\psi\pi^+\pi^-$ mass within $\pm$20\,MeV of the known $\Bsb$ mass and within $\pm$90\,MeV of the nominal $f_0(980)$ mass, are shown in Fig.~\ref{hel-980}; the data have been background subtracted, using the like-sign data, and acceptance corrected using Monte Carlo simulation. We perform a two-dimensional unbinned angular fit.  The ratio of rates is found to be
\begin{eqnarray}
\frac{\left| A_{20}\right|^2}{\left| A_{00}\right|^2}~~~~~&=&(0.1^{+ 2.6}_{-0.1})\%, \nonumber\\
\frac{\left| A_{21}\right|^2+\left| A_{2-1}\right|^2}{\left| A_{00}\right|^2}&=&(0.0^{+1.7}_{-0.0})\%,
\end{eqnarray}
where the uncertainties are statistical only.
The spin-2 amplitudes are consistent with zero. Note that the  $A_{20}$ amplitude corresponds to \CP odd final states, and thus would exhibit the same \CP violating phase as the $J/\psi f_0$ final state, while the $A_{2\pm1}$ amplitude can be either \CP odd or even. Thus this sample is taken as pure \CP odd.

%Considering the two-dimensional fit the $\chi^2$/ndof in 10x20 bins is 205/196 for both S and D-wave compared with 208/199 for only S-wave.  For the one-dimension distributions, the S-wave only fit, the $\chi^2$/ndof are 19.0/19 for the $\cos\theta_{J/\psi}$ distribution and 14.3/9 for $\cos\theta_{f_0}$. Adding in the D-wave changes the $\chi^2$/ndof to 12.8/6, which corresponds to a p-value of 4.6\% compared to S-wave only which is 11.3\%.  

\section{Time resolution and acceptance}
\label{sec:time}
The $\Bsb$ decay time is defined here as
$t^{\rm} = m \, {\vec{d}\cdot\vec{p}}/{|\vec{p}|^2}$,
where $m$ is the reconstructed invariant mass, $\vec{p}$ the
momentum and $\vec{d}$ the flight vector of the
candidate $\Bsb$ from the primary to the secondary vertices. If more
than one primary vertex is found, the one that corresponds to the
smallest IP $\chi^2$
of the $\Bsb$ candidate is chosen.

\begin{figure}[hbt]
\centering
\includegraphics[width=4.5in]{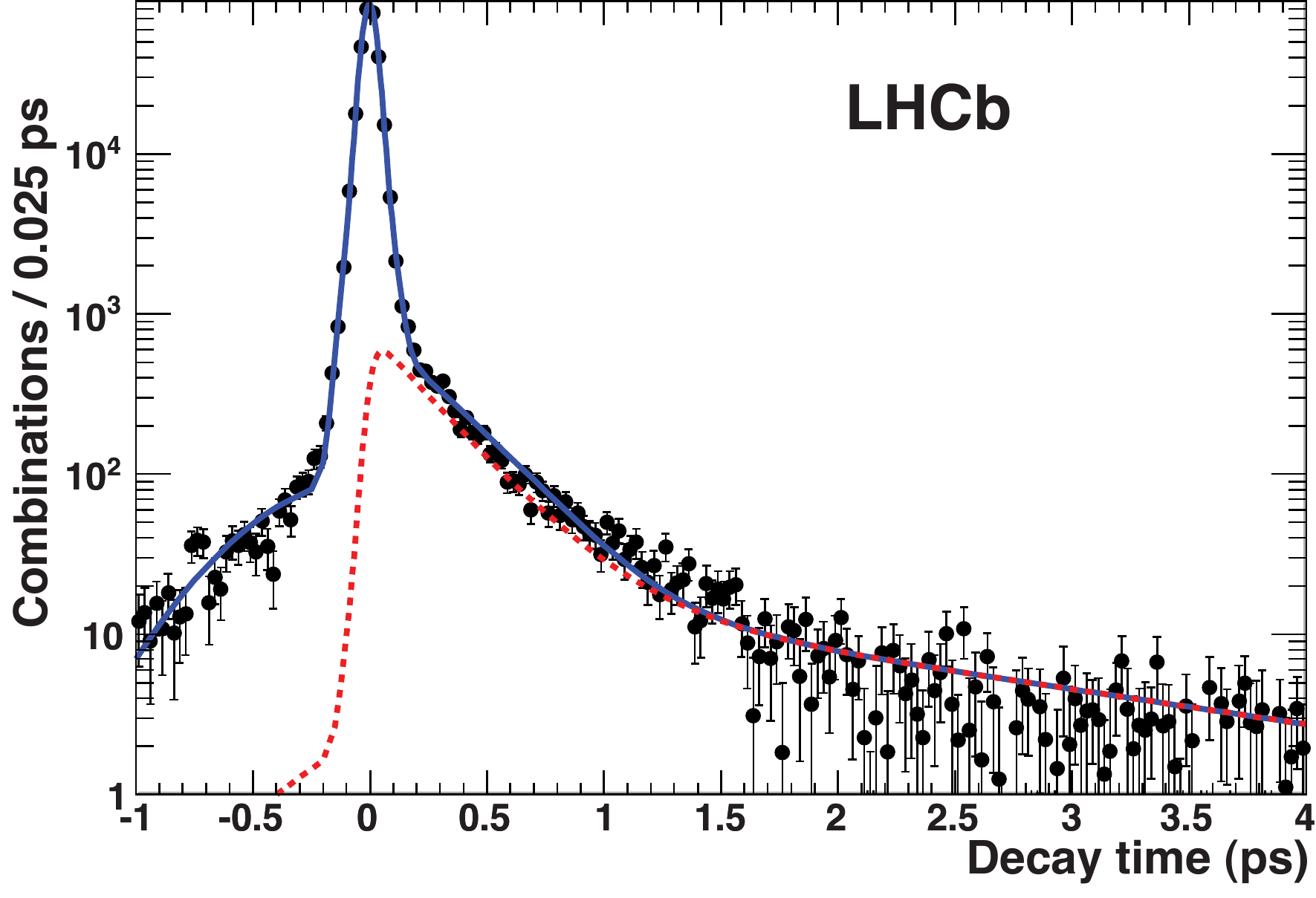}
\caption{Decay time distribution for prompt $J/\psi\pi^+\pi^-$ events. The dashed line (red) shows the long lived components, while the solid line (blue) shows the total.} \label{timeresDATA}
\end{figure}

The decay time resolution probability distribution function (PDF) is determined from data using $J/\psi$ detected without any requirement on detachment from the primary vertex (prompt) plus two oppositely charged particles from the primary vertex with the same selection criteria as for $J/\psi f_0$ events, except for 
the IP $\chi^2$ requirement. Monte Carlo simulation shows that the time resolution PDF is well modelled by these events. 
Fig.~\ref{timeresDATA} shows the $t^{\rm}$ distribution for our $J/\psi\pi^+\pi^-$ prompt 2011 data sample. To describe the background time distribution three components are needed, (i) prompt, (ii) a small long lived background ($f_{\rm LL1}= 2.64\pm 0.10$)\% modeled by an exponential decay function, and (iii) an even smaller component ($f_{\rm LL2}=0.46\pm 0.02$)\% from $b$-hadron decay described by an additional exponential. Each of these are convolved individually with a triple-Gaussian resolution function with common means, whose components are listed in Table~\ref{PDFs}. The overall equivalent time resolution is $\sigma_t$= 38.4 fs. 
%It is computed by defining 
%\begin{equation}
%{\cal D}=\sum_j f_j \exp\left(-\Delta m_s^2\sigma_{j}^2/2\right),
%\end{equation}
%where the fractions $f_j$,  and corresponding time resolutions $\sigma_j$ can be found in Table~\ref{PDFs}.
%Then 
%\begin{equation}
%\sigma_t =\sqrt{\left(-2\ln({\cal D}\right)}/\Delta m_s .
%\end{equation}

The functional form for the time dependence is given by 
\begin{eqnarray}
N(t)&=&(1-f_{\rm LL1}-f_{\rm LL2})\cdot 3G
+ f_{\rm LL1}\left[\frac{1}{\tau_1} \exp(-t/\tau_1)\otimes 3G\right]  \nonumber\\
&&+f_{\rm LL2}\cdot\left[1/\tau_2\cdot\exp(-t/\tau_2)\otimes 3G\right].
\end{eqnarray}
The fractions $f_{\rm LL1}$ and $f_{\rm LL2}$ , and their respective lifetimes $\tau_1$ and $\tau_2$,  are varied in the fit. The parameters of the triple-Gaussian time resolution, $3G$,  are listed in Table~\ref{PDFs}. The symbol $\otimes$ indicates a convolution.

A decay time acceptance is introduced by the triggering and event selection requirements.
Monte Carlo simulations show that the shape of the decay time acceptance function is well modelled by
\begin{equation}
\label{eq:A}
A(t)=C\frac{\left[a\left(t-t_0\right)\right]^n}{1+\left[a\left(t-t_0\right)\right]^n}~~,
\end{equation}
where $C$ is a normalization constant. Furthermore, the parameter values are found to be the same for simulated $\overline{B}^0\to J/\psi \overline{K}^{*0}$ events with $\overline{K}^{*0}\to K^-\pi^+$, as for $\Bsb\to J/\psi f_0$.
%\begin{figure}[hbt]
%\centering
%\includegraphics[width=6in]{accept-t}
%\caption{Time acceptance distributions for simulated (a) $J/\psi f_0$ events  and (b) $J/\psi \overline{K}^{*0}$ events. Note that the correlation between between $t_0$ and $n$ is $-0.816$. } \label{Taccept}
%\end{figure}
%he parameters are consistent within uncertainties showing the efficacy of using the  $J/\psi \overline{K}^{*0}$ events.

\begin{figure}[hbt]
\centering
\includegraphics[width=6.in]{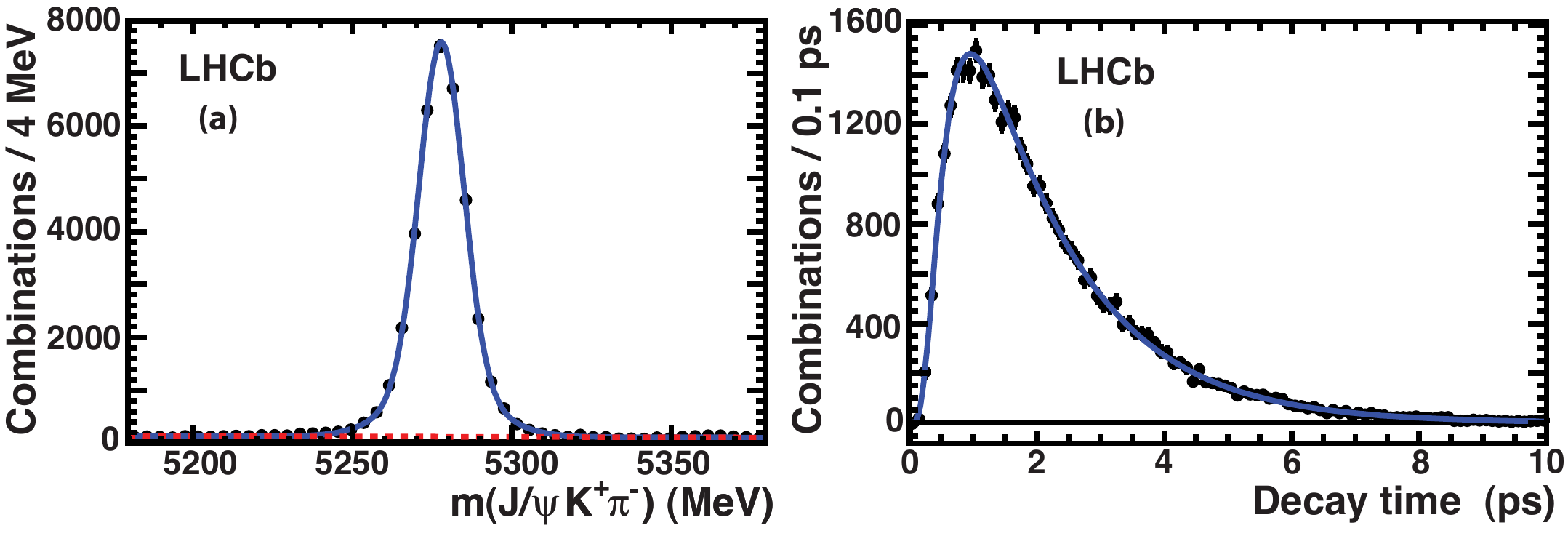}
\caption{Distributions for  $\overline{B}^0\to J/\psi \overline{K}^{*0}$ events (a) $\overline{B}^0$ candidate mass distribution  and (b) decay time distribution, where the small background has been subtracted using the $\overline{B}^0$ candidate mass sidebands.} \label{Jpsikstar}
\end{figure}

Fig.~\ref{Jpsikstar}(a) shows the $ J/\psi \overline{K}^{*0}$ mass distribution in data with  an additional requirement that the kaon candidate be positively identified in the RICH system, and that the $K^-\pi^+$ invariant mass be within $\pm$100\,MeV of 892 MeV. There are 36881$\pm$208 signal events. The sideband subtracted decay time distribution is shown in Fig.~\ref{Jpsikstar}(b) and fit using the above defined acceptance function gives values of $a=(1.89\pm 0.07)$\,ps$^{-1}$, $n=1.84\pm 0.12$, $t_0=(0.127\pm 0.015)$\,ps , and also a value of the $\overline{B}^0$ lifetime of 1.510$\pm$0.016\,ps, where the error is statistical only.  This is in good agreement with the PDG average of 1.519$\pm$0.007\,ps \cite{PDG}. 

%Another check is provided by a recent CDF measurement where they fit their $\Bsb\to J/\psi f_0$ data to a single exponential to extract an effective lifetime, $\tau^{\rm eff}$ \cite{Aaltonen:2011nk}. While this number only gives the lifetime of the lower mass eigenstate if $\phi_s$ equals zero (see Eqs.~\ref{eq:untagged}), we can fit our data and see if we have agreement. CDF measures 
%$\tau^{\rm eff}=1.70^{+0.12}_{-0.11}\pm 0.03$\,ps, where the first uncertainty is statistical and the second systematic. Our fit to a single exponential yields $\tau^{\rm eff}=1.68\pm 0.05$\,ps, again the error is only statistical.

Another check is provided by a recent CDF lifetime measurement of  $\Bsb\to J/\psi f_0$ of $1.70^{+0.12}_{-0.11}\pm 0.03$\,ps obtained by fitting the data to a single exponential \cite{Aaltonen:2011nk}. Such a fit to our data yields $1.68\pm 0.05$\,ps, where the uncertainty is only statistical.
\section{Fit strategy}

\subsection{Likelihood function characterization}
 The selected events are used to maximize a likelihood function
\begin{equation}
{\cal L}=\prod_i^{N} P(m_i,t_i^{\rm}, q_i),
\end{equation} 
where $m_i$ is the reconstructed candidate $\Bsb$ mass, $t_i^{\rm}$ the decay time, and $N$ the total number of events. The flavour tag, $q_i$, takes values of +1, $-1$ and 0, respectively, if the signal meson is tagged as $B_s^0$,  $\Bsb$, or untagged. The likelihood contains three components: signal, long-lived (LL) background and short-lived (SL) background.

For tagged events we have 
\begin{eqnarray}
P(m_i, t_i^{\rm}, q_i)&=&N_{\rm sig}\epsilon_{\rm sig}^{\rm tag} P_m^{\rm sig}(m_i)P_t^{\rm sig}(t_i^{\rm},q_i)\nonumber\\
&&+N_{\rm LL}\epsilon_{\rm LL}^{\rm tag}P_m^{\rm bkg}(m_i)P_t^{\rm LL}(t_i^{\rm})
+N_{\rm SL}\epsilon_{\rm SL}^{\rm tag}P_m^{\rm bkg}(m_i)P_t^{\rm SL}(t_i^{\rm}), 
\end{eqnarray}
where: (i) $P_m^{\rm sig}(m_i)$ and $P_m^{\rm bkg}(m_i)$
are the PDFs describing the dependence on reconstructed mass $m_i$ for signal and background events; (ii)
$P_t^{\rm sig}(t_i^{\rm},q_i)$ is the PDF used to describe
the signal decay rates for the decay time $t_i^{\rm}$; (iii)
$P_t^{\rm LL}(t_i^{\rm})$ is the PDF describing the long-lived
background decay rates,  and  $P_t^{\rm SL}(t_i^{\rm})$ describes the short-lived  background, both of which do not depend on the tagging; (iv)
$\epsilon^{\rm tag}$ refers to the respective tagging efficiencies for signal, long-lived and short-lived backgrounds.

For untagged events we have
\begin{eqnarray}
P(m_i, t_i^{\rm},0)&&=N_{\rm sig}(1-\epsilon_{\rm sig}^{\rm tag}) P_m^{\rm sig}(m_i)P_t^{\rm sig}(t_i,0)\nonumber\\
&&+N_{\rm LL}(1-\epsilon_{\rm LL}^{\rm tag})P_m^{\rm bkg}(m_i)P_t^{\rm LL}(t_i^{\rm})
+N_{\rm SL}(1-\epsilon_{\rm SL}^{\rm tag})P_m^{\rm bkg}(m_i)P_t^{\rm SL}(t_i^{\rm}).
\label{eq:Psig}
\end{eqnarray}

The total yields of the signal and background components are
fixed  to the number of events determined from the fit to the mass distributions (see Sec.~\ref{sec:2}). For both, the PDF is a product which models the invariant mass distribution and the
time-dependent decay rates. The $\Bsb$
mass spectrum is described by a double-Gaussian for the signal and an exponential
function for the background (see Fig.~\ref{fitmass_f0}). 
From Eqs. \ref{eq:CPrate} and \ref{eq:untagged}, the
decay time function for the signal is
\begin{equation}
\label{eq:R}
R(t,q_i) \propto e^{-\Gamma_s t}\left\{\cosh\frac{\Delta\Gamma_s
t}{2}+\cos\phi_s\sinh\frac{\Delta\Gamma_s t}{2}- q_i D
\sin\phi_s\sin(\Delta m_s t)\right\}.
\end{equation}
The probability of a wrong tag, $\omega$, is included in
the dilution factor $D\equiv(1-2\omega)$ (see Section~\ref{sec:flavortag}).   
%This function is smeared by time resolution then multiplied by the decay time acceptance function.

The signal PDF is taken as a product of the decay time function, $R(t,q_i)$, convolved with the triple Gaussian time resolution function multiplied with the time acceptance function found from $J/\psi K^{*0}$ discussed in Section~\ref{sec:time}.
The background decay time PDFs are determined using the like-sign $\pi^{\pm}\pi^{\pm}$ combinations. The time distribution of the like-sign background agrees in both yield and shape with the opposite-sign events in the upper $\Bsb$ mass candidate sideband 50$-$200\,MeV above the mass peak.

%The time distribution for like-sign events in the $f_0$ signal peak is shown in Fig.~\ref{fit_t_bkg}. 
%\begin{figure}[hbt]
%\centering
%\includegraphics[width=5.in]{fit_t_bkg}
%\caption{The decay time distribution of like-sign events fit to the combination of PDF's for both long-lived and short-lived backgrounds. The discontinuity of the fitted curve at very small decay time is due to short lived background.} \label{fit_t_bkg}
%\end{figure}
The background functions and parameters are listed in Table~\ref{PDFs}. The short-lived background component results from combining prompt $J/\psi$ events with a opposite-sign pion pair that is not rejected by our selection requirements. The long-lived part constitutes $\approx$85\% of the background.

\renewcommand{\arraystretch}{1.1}

\begin{table}[htb]
\center
\caption{\label{PDFs} The PDFs for the invariant mass and proper
time describing the signal and background. $P_t^{\rm sig}$ refers to the decay time distribution in Eq.~\ref{eq:Psig} and $A$ is given in Eq.~\ref{eq:A}. Where two numbers are listed, the first refers to the 2011 data and the second to the 2010 data. If only one number is listed they are the same for both years. The symbol $\hat{t}$ refers to the true time.}
\vspace{1mm}
\begin{tabular}{cc|l}\hline\hline
&$P_m$ & $P_t$ \\\hline 
\multicolumn{3}{c}{Signal~~~~~~~~~~~~~~~~~~~~~~~~~~~~~~~~}\\\hline
&Double-Gaussian ($2G$)&$P_t^{\rm sig}(t,q)=R(\hat{t},q)\otimes 3G(t^{\rm}-\hat{t};\mu,\sigma_1^t,\sigma_2^t,\sigma_3^t,f_2^t,f_3^t)$
\\
&$2G(m;m_0,\sigma_1,\sigma_2, f_2)$&$\cdot A(t^{\rm};a,n,t_0)$ \\
&$m_0$= 5366.5(3) MeV&$\mu=-0.0021(1)$\,ps, $-0.0011(1)$\,ps\\
&$\sigma_1$=8.6(3) MeV~~~~~~~&$\sigma_1^t=0.0300(4)$\,ps, $0.0295(5)$\,ps\\
&$\sigma_2$=26.8(9) MeV~~~~~&$\sigma_2^t/\sigma_1^t=1.92(4)$, $1.88(3)$ \\
&$f_2 $= 0.14(2)~~~~~~~~~~~&$\sigma_3^t/\sigma_1^t=14.6(10)$, $14.0(9)$ \\
&&$f_2^t=0.23(2)$, $0.27(3)$\\
&&$f_3^t=0.0136(6)$, $0.0121(7)$\\
&&$a=1.89(7)$\,ps$^{-1}$, $n=1.84(12)$, $t_0=0.127(15)$\,ps\\\hline
\multicolumn{3}{c}{Long-lived background~~~~~~~~~~}\\\hline
&Exponential&$[e^{-\hat{t}/{\tau^{\rm bkg}}}\otimes 2G(t^{\rm}-\hat{t};\mu,\sigma_1^t,\sigma_2^t,f_2^t)]\cdot
A(t^{\rm};a,n,t_0)$\\
&&$\mu=0$\\
&&$\sigma_1^t=0.088$\,ps\\
&&$\sigma_2^t=5.94$\,ps\\
&&$f_2^t=0.0137$\\
&&$\tau^{\rm bkg}=0.96$\,ps\\
&& $a=4.44$\,ps$^{-1}$, $n=4.56$, $t_0=0$\,ps\\
\hline
\multicolumn{3}{c}{Short-lived background~~~~~~~~~~}\\\hline
&Exponential&$2G(t^{\rm};\mu,\sigma_1^t,\sigma_2^t,f_2^t)\cdot
A(t^{\rm};a,n,t_0)$\\
&&All parameters are the same as for LL background\\
\hline\hline
\end{tabular}
\end{table}
\renewcommand{\arraystretch}{1.0}
\clearpage

\subsection{Flavour tagging}
\label{sec:flavortag}
%The procedure adopted here follows the work done in the analysis of $\Bsb\to J/\psi\phi$ decays \cite{LHCb-PAPER-2011-021}. 
Flavour tagging uses decays of the other $b$ hadron in the event, exploiting information from several sources including high transverse momentum muons, electrons and kaons, and the charge of  inclusively reconstructed secondary vertices.  The decisions of the four tagging algorithms are individually calibrated using $B^-\to  J/\psi K^-$ decays and combined \cite{LHCb-PAPER-2011-027}. 
The effective tagging performance is characterized by $\epsilon^{\rm tag}_{\rm sig} D^2$, where $\epsilon^{\rm tag}_{\rm sig}$ is the efficiency and $D$ the dilution.
% where $D$ is the dilution given in terms of the wrong sign of the flavour tag (mistag) fraction $\omega$ as $D=1-2\omega$. 
 We use a  per-candidate analysis that uses both the information of the tag decision and of the predicted mistag probability to classify and assign a weight to each event. The PDFs of the predicted mistag are taken from the  side-bands for the background and side-band subtracted data for the signal.

The calibration procedure uses a linear dependence between the estimated per event mistag probability $\eta$ and the actual mistag probability $\omega$ given by
$\omega=p_0+p_1\cdot\left(\eta-\langle \eta\rangle \right)$,
where $p_0$ and $p_1$ are calibration parameters and $\langle \eta \rangle$ is the average estimated mistag probability as determined from the calibration sample.
% With this parameterization there is essentially no correlation between $p_0$ and $p_1$.
%The values of these parameters are 
%given in Table~\ref{tab:ft} for both the 2010 and 2011 data samples. 
In the 2011 data $p_0=0.384\pm 0.003\pm 0.009$,  $p_1=1.037\pm0.040\pm0.070$, and  $\langle \eta \rangle=0.379$, with similar values in the 2010 sample. In this paper whenever two errors are given, the first is statistical and the second systematic.
Systematic uncertainties are evaluated by using different channels to perform the calibration including $\overline{B}^0\to D^{*+}\mu^-\overline{\nu}$, $B^+\to J/\psi K^+$ separately from $B^-\to J/\psi K^-$, and viewing the dependence on different data taking periods.
For our 2011 sample $\epsilon^{\rm tag}_{\rm sig}$ is (25.6$\pm$1.3)\% providing us with 365$\pm$22 tagged signal events.  For signal the mean mistag fraction, $\langle\eta\rangle$, is 0.375$\pm$0.005, while for background the mean is 0.388$\pm$0.006.
After subtracting background using like-sign events, we determine $D=0.289$
leading to an $\epsilon D^2$ of 2.1\%  \cite{LHCb-PAPER-2011-027}.

\section{Results}
Several parameters are input as Gaussian constraints in the fit. These include the
 LHCb measured value of $\Delta m_s =(17.63\pm 0.11\pm 0.02)$\,ps$^{-1}$ \cite{Aaij:2011qx}, the tagging parameters $p_0$ and $p_1$, and both the decay width given by the $J/\psi\phi$ analysis of $\Gamma_s=(0.657\pm0.009\pm0.008)$\,ps$^{-1}$ and  $\Delta\Gamma_s=(0.123\pm 0.029\pm 0.011$)\,ps$^{-1}$  \cite{LHCb:2011aa}; we also include the correlation of $-0.30$ between $\Gamma_s$ and  $\Delta\Gamma_s$.\footnote{The final fitted values of these parameters are shifted by less than 2\% from their input values.} The fit has been validated both with samples generated from PDFs and with full Monte Carlo simulations.
 
\begin{figure}[hbt]
\centering
\includegraphics[width=4.5in]{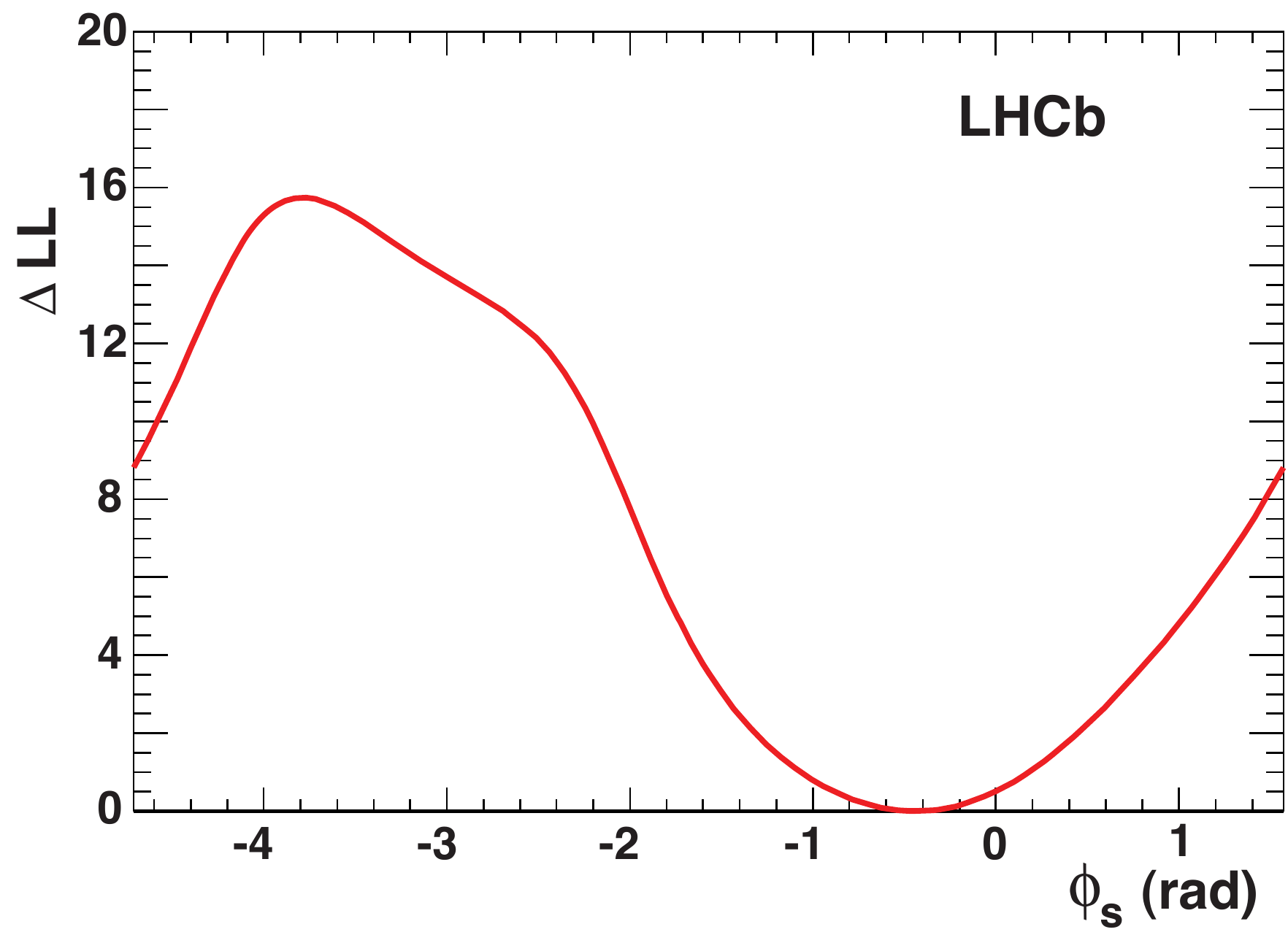}
\vspace*{-4mm}
\caption{Log-likelihood profile of $\phi_s$  for $\Bsb\to J/\psi f_0$ events.} \label{dll-new}
\vspace*{4mm}
\end{figure}

\begin{figure}[hbt]
\centering
\includegraphics[width=4.5in]{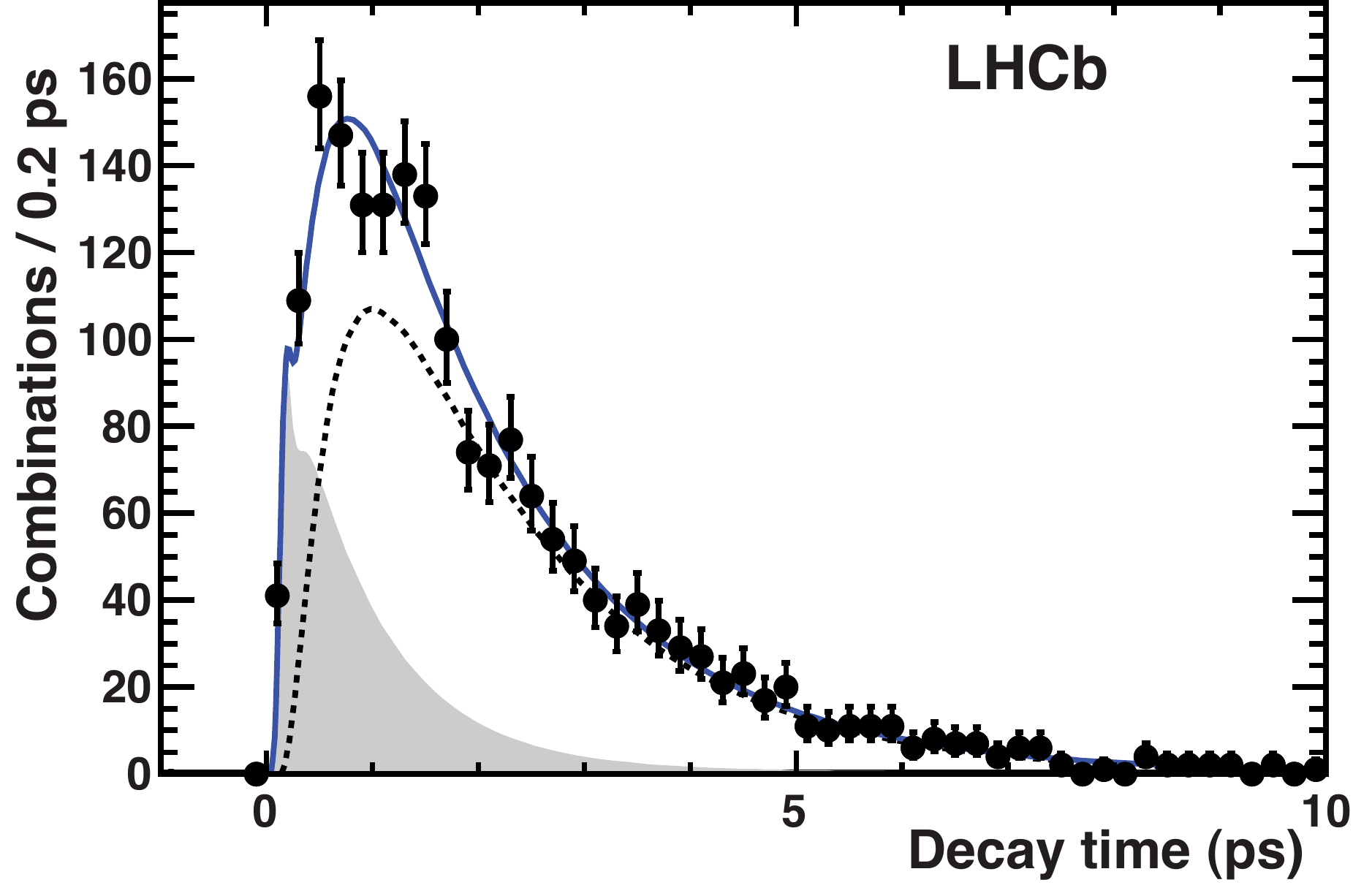}
\caption{ Decay time distribution from the fit for $J/\psi f_0$ candidates. The solid line shows the results of the fit, the dashed line shows the signal, and
the shaded region the background.}
%The discontinuity of the fitted curve at very small decay time is due to short lived background.}
 \label{time}
\end{figure}

Fig.~\ref{dll-new} shows the difference of log-likelihood value compared to that at the point with the best fit, as a function of $\phi_s$. At each $\phi_s$ value, the likelihood function is maximized with respect to all other parameters.  
The best fit value is $\phi_s= -0.44\pm 0.44$\,rad. The projected decay time distribution is shown in Fig.~\ref{time}.

\section{Systematic uncertainties}

The systematic errors are small compared to the statistical errors. No additional uncertainty is needed for  errors on  $\Delta m_s$, $\Gamma_s$, $\Delta\Gamma_s$ or flavour tagging, since Gaussian constraints are applied in the fit.  Other uncertainties associated parameters fixed in the fit are evaluated by changing them by $\pm$1 standard deviation from their nominal values and determining the change in fit value of $\phi_s$. These are listed in Table~\ref{tab:syserr}. An additional uncertainty is included due to the possible  \CP even D-wave. This has been measured at $(0.0^{+1.7}_{-0.0})$\% of the S-wave and contributes a small error to $\phi_s$, +0.007\,rad, as determined by repeating the fit with the mistag rate increased by 1.7\%. The asymmetry in production between $B^0_s$ and $\Bsb$ is believed to be small, about 1\%, and similar to the same asymmetry in $B^0$ production which has been measured by LHCb to be about 1\% \cite{LHCb-CONF-2011-042}. The effect of neglecting a 1\% production asymmetry is the same as ignoring a 1\% difference in the mistag rate and causes negligible bias in $\phi_s$.
%The systematic uncertainties are small compared with the statistical error and are ignorable.
\begin{table}[!htb]
\centering
\caption{Summary of systematic uncertainties. Here $N_{\rm bkg}$ refers to the number of background events, $N_{\rm sig}$ the number of signal,  $N_{\eta'}$ the number of $\eta'$, $\alpha$ the exponential background parameter for the $\Bsb$ candidate mass, $N_{\rm LL}/N_{\rm bkg}$ 
the long-lived background fraction. The Gaussian signal parameters are the mean $m_0$,
the width  $\sigma(m)$; $t_0$, $a$ and $n$ are the three parameters in the acceptance time function. The resolution in signal time is given by $\sigma(t)$, and the background lifetime by $\tau_{\rm bkg}$.
The final uncertainty is found by adding all the sources in quadrature.}
\label{tab:syserr}
\begin{tabular}{lcrr}
\hline\hline
Quantity (Q) & $\pm\Delta$Q & $+$Change& $-$Change  \\
              &                                    &in $\phi_s~~$ & in $\phi_s$~~\\
\hline
$N_{\rm bkg}$  & 10.1 &0.0025 &	$-0.0030$\\
$N_{\eta'}$ & 3.4	&$-0.0001$& $-0.0001$\\	
$N_{\rm sig}$ &46.47&$-0.0030$& 0.0028\\
$\alpha$ &$1.7\cdot 10^{-4}$&$-0.0002$&$-0.0002$\\
$N_{\rm LL}/N_{\rm bkg}$ &0.0238&0.0060&$-0.0063$\\
$m_0$ (MeV)& 0.32&-0.0003&	0.0011\\
$\sigma(m)$ (MeV)& 0.31& $-0.0026$&0.0020\\
$\tau_{\rm bkg}$ (ps) &0.05&$-0.0075$&0.0087\\
$\sigma(t)$ (ps) &5\%& $-0.0024$&0.0022\\
$t_0$ (ps) & 0.015&$0.0060$&0.0050\\
$a$ (ps$^{-1}$)& 0.07&$-0.0065$&$-0.0065$\\
$n$&0.12&	$-0.0089$&$-0.0089$\\
\CP-even D-wave && $0.0070$ & 0\\
\hline
\multicolumn{2}{l}{Total Systematic Error} &+0.018 &$-0.017$\\
\hline\hline
\end{tabular}
\end{table}

\section{Conclusions} 
Using 0.41\,fb$^{-1}$ of data collected with the LHCb detector, the decay mode
$\Bsb\to J/\psi f_0$, $f_0\to\pi^+\pi^-$ is selected and then used to measure the \CP violating phase, $\phi_s$. 
We perform a time dependent fit of the data with the $\Bsb$ lifetime  and the difference in widths of the heavy and light eigenstates constrained. Based on the likelihood curve in Fig.~\ref{dll-new} we find 
\begin{equation}
\phi_s= -0.44\pm 0.44\pm0.02~{\rm\,rad}, \nonumber
\end{equation}
consistent with the SM value of $-0.0363^{+0.0016} _{-0.0015}$ \,rad  \cite{Charles:2011va}.  Assuming the SM , the probability to observe our measured value is 36\%.
%The  probability that our result is consistent with the Standard Model is 36\%. 
There is an ambiguous solution with $\phi_s\to\pi-\phi_s$ and $\Delta\Gamma_s\to -\Delta\Gamma_s$.
The precision of the result mostly results from using the tagged sample, though the untagged events 
also contribute.

LHCb provides an independent measurement of $\phi_s=0.15\pm 0.18\pm 0.06$ \cite{LHCb:2011aa} using the $\Bsb\to J/\psi\phi$ decay. Combining these two results, taking into account all correlations by performing a joint fit, we obtain
\begin{equation}
\phi_s=0.07\pm0.17\pm0.06~{\rm rad}~~{\rm (combined)}.\nonumber
\end{equation}
This is the most accurate determination of $\phi_s$ to date, and is consistent with the SM prediction.

\section*{Acknowledgements}

\noindent We express our gratitude to our colleagues in the CERN accelerator
departments for the excellent performance of the LHC. We thank the
technical and administrative staff at CERN and at the LHCb institutes,
and acknowledge support from the National Agencies: CAPES, CNPq,
FAPERJ and FINEP (Brazil); CERN; NSFC (China); CNRS/IN2P3 (France);
BMBF, DFG, HGF and MPG (Germany); SFI (Ireland); INFN (Italy); FOM and
NWO (the Netherlands); SCSR (Poland); ANCS (Romania); MinES of Russia and
Rosatom (Russia); MICINN, XuntaGal and GENCAT (Spain); SNSF and SER
(Switzerland); NAS Ukraine (Ukraine); STFC (United Kingdom); NSF
(USA). We also acknowledge the support received from the ERC under FP7
and the Region Auvergne.

\clearpage
\newpage
\bibliographystyle{LHCb}
\bibliography{f0-1}
\end{document}